\documentclass[12pt]{article}
\usepackage{authblk}
\usepackage{tikz}
\usetikzlibrary{backgrounds}
\usepackage{bbm}
\usepackage{apacite}
\usepackage{natbib}
\usepackage{amssymb,amsmath, amsfonts}
\usepackage{hyperref}
\usepackage{amsmath}
\usepackage{subfig}
\usepackage{graphicx}
\usepackage{placeins}
\usepackage{booktabs}
\usepackage{rotating}
\usepackage[english]{babel}
\usepackage{epstopdf}
\usepackage{multirow}
\usepackage{array}
\usepackage{mathtools}
\usepackage[document]{ragged2e}
\usepackage{threeparttable}
\usepackage{caption}
\usepackage{color}
\usepackage[normalem]{ulem}
\usepackage{float}
\DeclareMathOperator{\Y}{\mathbf{Y}}
\DeclareMathOperator{\y}{\mathbf{y}}

\newcolumntype{R}[1]{>{\raggedleft\arraybackslash}p{#1}}

\begin{document}
	
\begin{titlepage}
\title{Cluster-weighted latent class modeling} 
 
\author[1]{Roberto Di Mari\thanks{roberto.dimari@unict.it}}
\author[1]{Antonio Punzo}
\author[2]{Zsuzsa Bakk}
\affil[1]{Department of Economics and Business, University of Catania, Italy}
\affil[2]{Leiden University, Institute of Psychology, Methodology \& Statistics Unit}

\date{\today}
\maketitle

\begin{abstract}
	Usually in  Latent Class Analysis (LCA), external predictors are taken to be cluster conditional probability predictors (LC models with covariates), and/or score conditional probability predictors (LC regression models). In such cases, their distribution is not of interest. Class specific distribution is of interest in the distal outcome model, when the distribution of the external variable(s) is assumed to dependent on LC membership. In this paper, we consider a more general formulation, typical in cluster-weighted models, which embeds both the latent class regression and the distal outcome models. This allows us to test simultaneously both whether the distribution of the covariate(s) differs across classes, and whether there are significant direct effects of the covariate(s) on the indicators, by including most of the information about the covariate(s) - latent variable relationship. We show the advantages of the proposed modeling approach through a set of population studies and an empirical application on assets ownership of Italian households. 
\end{abstract}
\end{titlepage}

\section{Introduction}\label{sec:intro}
Latent class analysis \citep{mccutcheon1985latent} is widely used in the social and behavioral sciences to locate subgroups of observations in the sample based on a set of $J$ observed response variables $\mathbf{Y}$. Examples of applications include identification of types of mobile internet usage in travel planning and execution \citep{okazaki15}, types of
political involvement \citep{hagenaars1989}, classes of treatment engagement in adolescents with psychiatric problems \citep{roedelof13}, a typology of infant temperament \citep{loken2004}, modeling phases in the development of transitive reasoning \citep{bouwmeester}, or classes of self disclosure \citep{henk}.

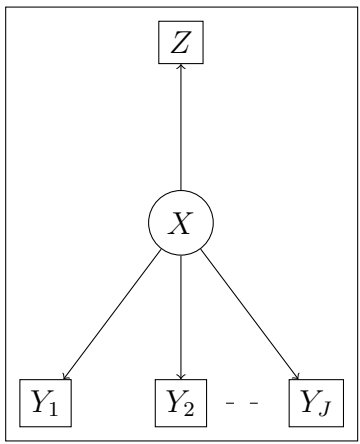
\begin{figure}[!h]
    \centering
\begin{tikzpicture}[framed,scale=0.6]
\tikzstyle{every node}=[draw,shape=circle,remember picture] %
\node[rectangle] (v0) at (6,6) {$Z$}; 
\node
(v1) at (6,2) {$X$}; 
\node[rectangle] (v2) at (3,-2)
{$Y_{1}$}; \node[rectangle] (v3) at (6,-2) {$Y_{2}$};
\node[rectangle] (v4) at (9,-2) {$Y_{J}$}; 
\draw[->] (v1) -- (v0); \draw[->] (v1) -- (v2); \draw[->] (v1) -- (v3);
\draw[->] (v1) -- (v4); 
\draw[loosely dashed] (7,-2) -- (8,-2);
\end{tikzpicture}
\caption{\footnotesize \label{fig:distalmod} Latent class model with distal outcome.}
\end{figure}

In many empirical studies, interest lies in investigating which external variables $\mathbf{Z}$ predict latent class membership $X$. Latent class models with covariates \citep{dayton} are a well-known extension of the baseline model, in which external variables are included in the latent class modeling framework as predictors of class membership \citep{collins2010latent}. \cite{stegmann2017}, for instance, discuss the inclusion of covariates in more complicated LC models. However, recent methodological development has shifted the attention towards modeling the effect in the opposite direction. That is, predicting a - possibly continuous - distal outcome based on the latent class membership \citep{bakk:12, lanza:13}, as depicted in Figure \ref{fig:distalmod}. Although there can be more than one external variable available, for sake of exposition here we describe the models with only one external variable\footnote{All considered modeling scenarios can be straightforwardly extended to the multiple external variables case. See, for instance, \cite{bakk:12}}.

For instance, \cite{roberts2011} predict distal pain outcomes based on class memberships defined by patterns of barriers to pain management and \cite{mulder2012} compared average measures of recidivism in clusters of juvenile offenders.

Typically, in distal outcome models, the distal outcome and the $J$ response variables $\mathbf{Y}$'s are assumed to be conditionally independent given the latent variable $X$ \citep{bakk:12, lanza:13}. A direct effect of $Z$ on $\mathbf{Y}$ is therefore not allowed for, neither its presence tested. 
In latent variable modeling, it is well known that Maximum Likelihood (ML) estimation is subject to severe bias when direct effects are present in LC and latent trait models (\citealp{asparouhov2012auxiliary}, regression mixture models \citep{leepaper,nylund16}, and latent Markov models \citep{robzsuzsa17}, and are not accounted for. %

Given the restrictiveness of the conditional independence assumption and the possible severity of its violation, we propose a more general model that can account for complex interdependencies between the external variable, LC membership, and the indicators of the LC model. 
In regression mixtures, a ``circular'' relation among $\mathbf{Y}$-$X$-$Z$ is commonly considered in the cluster-weighted modeling approach \citep{ingrassia2012,Ingr:Mino:Punz:Mode:2014,Ingr:Punz:Vitt:TheG:2015,Punz:Flex:2014,Dang:Punz:McNi:Ingr:Brow:Mult:2017}. That is, a more general model is specified, where next to modeling the class specific distribution of $Z$ (distal outcome situation), also the direct effect of $Z$ on $\mathbf{Y}$ is modeled (latent class regression). If $\mathbf{Y}$ are indicators of assets ownership, and $Z$ is a measure (in euro) of net (of liabilities) wealth, the cluster-weighted modeling approach allows net wealth also to directly affect a household decision to own assets. With standard inference, the statistical significance of each effect can then be tested to see whether intermediate model specifications are more appropriate.

\begin{figure}[!h]
    \centering
\begin{tikzpicture}[framed,scale=0.5]
\tikzstyle{every node}=[draw,shape=circle,remember picture] %
\node[rectangle] (v0) at (6,6) {$Z$}; 
\node
(v1) at (6,2) {$X$}; 
\node[rectangle] (v2) at (3,-2)
{$Y_{1}$}; \node[rectangle] (v3) at (6,-2) {$Y_{2}$};
\node[rectangle] (v4) at (9,-2) {$Y_{J}$}; 
\node[rectangle] (v5) at (6,-6) {$Z$};
\draw[->] (v1) -- (v0); \draw[->] (v1) -- (v2); \draw[->] (v1) -- (v3);
\draw[->] (v1) -- (v4); \draw[->] (v5) -- (v2); \draw[->] (v5) -- (v3);
\draw[->] (v5) -- (v4);
\draw[loosely dashed] (7,-2) -- (8,-2);
\end{tikzpicture}
\caption{\footnotesize \label{fig:cwmlca} Latent class cluster-weighted model.}
\end{figure}
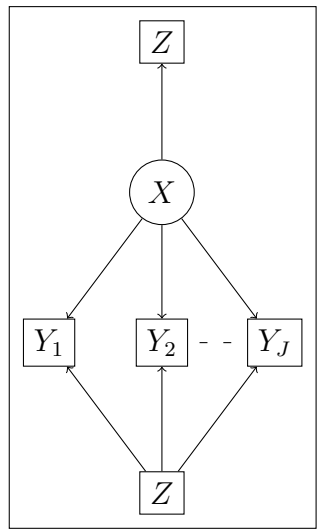

%
%
In LC regression models \citep{kamakura1989,wedel94}, although the assumption of conditional independence of $\mathbf{Y}$ and $Z$ can be relaxed (see Figure \ref{fig:latregmod}), the distal outcome's distribution is not of interest and hence not modeled. Therefore, in the traditional LCA approach, an external variable enters the model either as a covariate (latent class regression) or as a distal outcome, but never as both at the same time. We propose to extend the idea of cluster-weighted modeling in the context of latent class analysis, by proposing a generalized version of the models in Figures \ref{fig:distalmod} and \ref{fig:latregmod}, as depicted in Figure \ref{fig:cwmlca}, which embeds them both.
\begin{figure}[!h]
    \centering
\begin{tikzpicture}[framed,scale=0.6]
\tikzstyle{every node}=[draw,shape=circle,remember picture] %
\node
(v0) at (6,6) {$X$}; 
\node[rectangle] (v1) at (3,2)
{$Y_{1}$}; \node[rectangle] (v2) at (6,2) {$Y_{2}$};
\node[rectangle] (v3) at (9,2) {$Y_{J}$}; 
\node[rectangle] (v4) at (6,-2) {$Z$};
\draw[->] (v0) -- (v1); \draw[->] (v0) -- (v2); \draw[->] (v0) -- (v3);
\draw[->] (v4) -- (v1); \draw[->] (v4) -- (v2); \draw[->] (v4) -- (v3);
\draw[loosely dashed] (7,2) -- (8,2);
\end{tikzpicture}
\caption{\footnotesize \label{fig:latregmod} Latent class regression model.}
\end{figure}
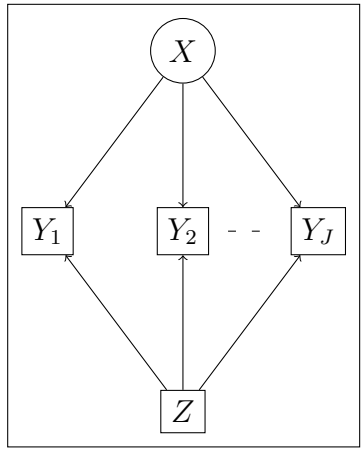

By starting from the most general model, the user can proceed backwards, testing the model assumptions of both the distal outcome and the latent class regression models. In particular, in this paper we will show evidence, based on a set of population studies and an empirical application, that 1) if direct effects are present, our approach, contrary to the distal outcome model, yields unbiased estimates of the distal outcome cluster specific means and variances; and 2) if the most suitable model is one between the distal outcome model or the latent class regression model, the relative class sizes and compositions will be the same as the ones delivered under the proposed modeling approach.

The paper proceeds as follows. 
In Section~\ref{sec:popstudy}, we illustrate the proposed modeling approach through three population studies, in comparison with the LC regression and the distal outcome models. 
We give model definitions and details on the parameterizations in Section~\ref{sec:models}. 
In Section~\ref{sec:empirical}, we analyze data from the Household Finance and Consumption Survey, and conclude with some final remarks in Section~\ref{sec:concl}.    

\section{Population studies}
\label{sec:popstudy} 

This Section is devoted to showing very simple and intuitive evidence, obtained by analyzing three large data sets (30000 sample units) - each drawn from the three models in Figures \ref{fig:distalmod}, \ref{fig:latregmod} and \ref{fig:cwmlca} - in order to motivate the application of the cluster-weighted modeling approach in LCA (see Table \ref{table:legend}). We set the number of latent classes $S = 2,$ and to begin with we fit all three models assuming this value to be known. At the end of the Section, we will also show results on estimation of the number of latent classes based on BIC. 
\begin{table}[!h]
\centering  
\begin{tabular}{lccc}
\hline \hline
									&	 & Acronym & Data \\
\hline
Latent Class regression 			&	&	LCreg	& \emph{LCreg} \\
Latent Class with distal outcome 	&	&	LCdist	& \emph{LCdist} \\
Latent Class cluster-weighted 		&	&	LCcw	& \emph{LCcw} \\
\hline \hline
\end{tabular}
\caption{\footnotesize Legend of acronyms used for the population models and for the generated data.\label{table:legend}}
\end{table}

To get approximately equal (realistic) conditions on class separation, we generated the data such that the entropy-based $R^2$ \citep{magidson81} for the correctly specified model is about 0.7 in all the three data sets - which is the minimum class separation to get a good LC model \citep{vermunt:10,asparouhov2012auxiliary}. The data were generated in R \citep{rcore}, and parameter estimation was carried out with Latent GOLD 5.1 \citep{latentG5.1}.

\subsection{\emph{LCreg} data}\label{sec:lcrdata}
The \emph{LCreg} data set was generated from a two-class LCreg model, with class memberships of 0.7 and 0.3, six dichotomous indicators $(J=6)$ and one continuous $Z$ - drawn from a standard normal distribution - loaded on all six indicators. The external variable $Z$ is loaded on the indicators with a coefficient of -0.5, if the most likely response is on the first class, or 1, if the most likely response is on the second class, giving a large effect size.

\begin{table}[!h]
\centering  
\begin{tabular}{lcccccc}
\hline \hline
& \multicolumn{2}{c}{Class proportions} & & Entr. $R^2$ & & \#par \\
\hline
LCreg &	{\bf 0.7010}&	{\bf 0.2990} 	& &{\bf 0.7675} & &{\bf 25}\\	
LCdist &	0.7357  &		0.2643  	& &    0.8639   & & 17   \\	
LCcw   &	0.7018	&		0.2982  	& &    0.7681   & & 29   \\	
\hline
\hline 
\end{tabular}
\caption{\footnotesize \emph{LCreg} data. Estimated class proportions, entropy-based $R^2$ and number of parameters for each of the three estimated models. Results from correctly specified model in bold font.\label{table:lcregclassprop}}
\end{table}
We observe in Table \ref{table:lcregclassprop} that the LCdist model overinflates the mixing proportion on the bigger class, whereas the LCcw model yields nearly equivalent class proportions as in the correctly specified case. This at the cost of four more parameters to be estimated. 

Table \ref{table:lcrmeanvar} reports estimated means and variances for the variable $Z$ based on the LCdist and LCcw models, along with standard errors and $p$-values of the Wald tests of equality of the means and the variances. Nothing is reported for LCreg, as $Z$ is not modeled. In the LCdist model, both the means are wrongly estimated to be statistically different from zero. Moreover, based on the reported Wald tests, we reject the nulls of equal means and equal variances (with $p$-values smaller than 0.01). These findings for the LCdist model can be explained by the fact that it wrongly predicts a clustered distribution on $Z$ in order to accommodate for a direct effect of $Z$ on the indicators which is not accounted for. This creates an additional source of entropy in the class solution (as displayed by the relatively higher value of the entropy-based $R^2$). 

\begin{table}[!h]
\centering  
\begin{tabular}{lllcllc}
\hline \hline
& \multicolumn{2}{c}{Means} & Wald(=) $p$ &  \multicolumn{2}{c}{Variances} &  Wald(=) $p$ \\
\hline
LCdist&		 0.0525*** &	     -0.1640***  &    0.0000	&  1.0301  &     0.8846  & 0.0000 \\	
&	(0.0071)		&	(0.0122)	&	  		&   (0.0100)   & (0.0158)     &     \\	
&					&				&			&   		   & 		     &     \\	
LCcw  &	-0.0010	&		-0.0134  &	0.9000  &     0.9966   & 1.0105       &  0.6100  \\	
&			(0.0086)  &     (0,0163)&			&    (0.0114)    &  (0.0208)     &     \\	
\hline
\hline 
\end{tabular}
\caption{\footnotesize \emph{LCreg} data. Estimated means (*** $p$-value$<$0.01, ** $p$-value$<$0.05, * $p$-value$<$0.1) and variances , and $p$-values from Wald test of  equality of component means and variances for the LCdist model and the LCcw model. Standard errors in parentheses.\label{table:lcrmeanvar}}
\end{table}

\subsection{\emph{LCdist} data}\label{sec:disdata}
The \emph{LCdist} data set was generated from a two-class LCdist model, with class memberships of 0.7 and 0.3, six dichotomous indicators ($J=6$) and one continuous $Z$, drawn from a mixture of two normal distributions with means of -1 and 1 and common variance of 1.
\begin{table}[!h]
\centering  
\begin{tabular}{lcccccc}
\hline \hline
& \multicolumn{2}{c}{Class proportions} & & Entr. $R^2$ & & \#par \\
\hline
LCreg &		0.5850  &		0.4150  	& &    0.2781   & & 25   \\	
LCdist&	{\bf 0.7006}&	{\bf 0.2994} 	& &{\bf 0.7274} & &{\bf 17}\\	
LCcw  &	0.7026	&		0.2974  	& &    0.7320   & & 29   \\	
\hline
\hline 
\end{tabular}
\caption{\footnotesize \emph{LCdist} data. Estimated class proportions, entropy $R^2$ and number of parameters for each of the three estimated models. Results from correctly specified model in bold font.\label{table:LCdistclassprop}}
\end{table}

The LCreg model yields a completely distorted class solution, whereby both the LCdist and LCcw models yields almost identical (correct) solutions (Table \ref{table:LCdistclassprop}). 

Interestingly, the misspecified response-$Z$ relation in the LCreg model yields a solution with relatively smaller class separation (as measured by the entropy-based $R^2$).

Next, we compare estimates of class-specific means and variances of $Z$ as obtained by the LCdist and LCcw models.

\begin{table}[!h]
\centering  
\begin{tabular}{lcccccc}
\hline \hline
& \multicolumn{2}{c}{Means} & Wald(=) $p$ &  \multicolumn{2}{c}{Variances} &  Wald(=) $p$ \\
\hline
LCdist&	{\bf -0.9911}***&	{\bf 1.0156}*** &  {\bf 0.0000}	& {\bf 1.0072} & {\bf 0.9886} & 0.4000 \\	
&	(0.0084)		&	(0.0145)	&	  		&   (0.0119)   & (0.0196)     &     \\	
&					&				&			&   		   & 		     &     \\	
LCcw  &	-0.9889***	&		1.0242***  &	0.0000  &     1.0075   & 0.9810       &  0.2700  \\	
&			(0.0096)  &     (0,0176)&			&    (0.0125)    &  (0.0207)     &     \\	
\hline
\hline 
\end{tabular}
\caption{\footnotesize \emph{LCdist} data. Estimated means (*** $p$-value$<$0.01, ** $p$-value$<$0.05, * $p$-value$<$0.1) and variances , and $p$-values from Wald test of equality of component means and variances for the LCdist model and the LCcw model. Standard errors in parentheses. Results from correctly specified model in bold font. \label{table:distmeanvar}}
\end{table}

The LCdist and the LCcw models estimate almost identical means and variances of $Z$, both correctly not rejecting the null of common variance across latent classes. We observe that the SE's for the less parsimonious LCcw model are systematically larger than those of the correctly specified model: this is not surprising, as having less degrees of freedom corresponds, all else equal, to slightly more variable estimates.

\subsection{\emph{LCcw} data}\label{sec:cwmdata}
The \emph{LCcw} data was generated from a two-class LCcw model, with class memberships of 0.7 and 0.3, six dichotomous indicators ($J=6$) and one continuous $Z$, drawn from a mixture of two normal distributions with means of -1 and 1 and common variance of 1.

\begin{table}[!h]
\centering  
\begin{tabular}{lcccccc}
\hline \hline
& \multicolumn{2}{c}{Class proportions} & & Entr. $R^2$ & & \#par \\
\hline
LCreg &	0.8899  &	0.1101  	& &    0.5611   & & 25   \\	
	 
LCdist  &	0.4373 	&		 0.5627 	& &    0.6441   & & 17   \\	
LCcw&	{\bf 0.6993}&	{\bf 0.3007} 	& &{\bf 0.7045} & &{\bf 29}\\
\hline
\hline 
\end{tabular}
\caption{\footnotesize \emph{LCcw} data. Estimated class proportions, entropy-based $R^2$ and number of parameters for each of the three estimated models. Results from correctly specified model in bold font.\label{table:lccwmclassprop}}
\end{table}

Both the LCreg and the LCdist models deliver distorted class solutions (Table \ref{table:lccwmclassprop}). Although with a higher entropy-based $R^2,$ the residual dependence among the indicators due to the exclusion of the direct effect causes a more severe distortion in the LCdist model compared to the LCreg model. Equally we observe (Table \ref{table:lccwmmeanvar}) that the means and variance(s) of $Z$ are both biased in the LCdist model. Contrary to the correctly specified model, in LCdist the Wald test cannot reject the equal variances hypothesis (at 1 \% level).
\begin{table}[!h]
\centering  
\begin{tabular}{lcccccc}
\hline \hline
& \multicolumn{2}{c}{Means} & Wald(=) $p$ &  \multicolumn{2}{c}{Variances} &  Wald(=) $p$ \\
\hline
LCdist&	 -1.4544***&	  0.4159*** &    0.0000 	&   0.7044 &    1.2096 & 0.0000 \\	
&	(0.0102)		&	(0.0120)	&	  		&   (0.0111)   & (0.0159)     &     \\	
&					&				&			&   		   & 		     &     \\	
LCcw  & {\bf -1.0029}***	&	{\bf 0.9955}***  &	{\bf 0.0000}  &   {\bf 1.0215}   & {\bf 0.9818}       &  {\bf 0.0380}  \\	
&			(0.0104)  &     (0,0122)&			&    (0.0146)  &     (0.0160)     &     \\	
\hline
\hline 
\end{tabular}
\caption{\footnotesize \emph{LCcw} data. Estimated means (*** $p$-value$<$0.01, ** $p$-value$<$0.05, * $p$-value$<$0.1) and variances , and $p$-values from Wald test of equality of component means and variances for the LCdist model and the LCcw model. Standard errors in parentheses. Results from correctly specified model in bold font. \label{table:lccwmmeanvar}}
\end{table}

Table \ref{table:ARIcomp} reports adjusted Rand indexes ARI \citep{hubertARI}, arranged in a three-by-three table, comparing the hard partitions obtained with each fitted model under the three data generating model scenarios. The results are in line with what observed above. When the data are generated with a LCreg model, the LCcw model delivers an almost identical partition to that of the correctly specified model, followed close up by the LCdist model - with only about 3\% difference. In the \emph{LCdist} data set as well, the LCcw model's partition is nearly as in the correctly specified model (ARI of $\approx 0.99$), whereby the ARI drops to $\approx 0.21$ when the comparison is with the LCreg partition. In the latest scenario - \emph{LCcw} data set - fitting both the LCreg and the LCdist models delivers in both cases quite different partitions ($\approx 0.16$ and $\approx 0.31$ ARI's) compared to the correctly specified model.    

\begin{table}[!h]
\centering  
\begin{tabular}{lclccc}
\hline \hline
& &	&\multicolumn{3}{c}{Fitted model}\\
\cmidrule{4-6}
Data & &Correct model	   &	LCreg	   &	LCdist		&    LCcwm	 	\\
\hline

\emph{LCreg}  & &LCreg  &	 1		   &	  0.9732	&    0.9997 	\\	
\emph{LCdist} & &LCdist &	 0.2125	   &	  1     	&    0.9889 	\\	
\emph{LCcw}  &  &LCcw  &	 0.1604    &	  0.3101    &    1      	\\	
\hline
\hline 
\end{tabular}
\caption{\footnotesize {\bf Adjusted Rand indexes}, computed between clustering with correctly specified models - LCreg, LCdist and LCcw models - and clustering with the other two models  \label{table:ARIcomp}}
\end{table}

Based on the above data sets, in Table \ref{table:BICcomp} we report also results on BIC values for the three models in all three scenarios, for $S=1,\dots,5$. Although BIC values can be compared for LCdist and LCcw, selecting a model among the three with BIC cannot be done as $Z$ in LCreg is not modeled and the model likelihoods are therefore not comparable. In both the \emph{LCreg} and \emph{LCdist} data sets, BIC for the LCcw model selects, together with the correctly specified model in the first two scenarios, the correct number of classes. Interestingly however, misspecifying the indicators-$Z$ relation causes, in both the LCreg and LCdist models, a severe overstatement of the number of classes (\emph{LCcw} data set).   

\begin{table}[!h]
\centering  
\begin{tabular}{lclccccc}
\hline \hline
&&&	\multicolumn{5}{c}{Number of components}\\
\cmidrule{4-8}
Data	&&	  			&	$S=1$   & $S=2$	   &	$S=3$		&    $S=4$	 	&     $S=5$\\
\hline

\multirow{3}{*}{\emph{LCreg} $\begin{dcases*} \\ \\ \end{dcases*}$}&&{\bf LCreg}  		& 235957.23 &{\bf 191309.94} &	191411.62	&  191539.73	& 191642.90		\\	
&&LCdist 				& 321109.39 & 283926.34  & 278042.32	&  277453.71	& {\bf 277117.53}\\	
&&LCcw  				& 321137.31 &{\bf 276509.33}  & 276636.54   &  276766.45	& 276888.98\\
\cmidrule{2-8}

\multirow{3}{*}{\emph{LCdist} $\begin{dcases*} \\ \\ \end{dcases*}$}&&LCreg  				& 233652.06 & 231508.46    &   231251.75 & 231041.07	& {\bf 230893.17}\\	
&&{\bf LCdist} 		& 348714.25	&	{\bf 331953.90} & 332037.14 & 332112.24 	& 332189.69\\	
&&LCcw  				& 337204.58 & {\bf 332067.91}    & 332198.43 & 332329.23	&	332449.70\\	
\cmidrule{2-8}

\multirow{3}{*}{\emph{LCcw } $\begin{dcases*} \\ \\ \end{dcases*}$} && LCreg  			   & 213395.81 &207054.91   &  206149.45	&  205996.43 	&	{\bf 205922.50}	\\	
&& LCdist 			   & 321792.35 & 310231.51   &	306431.47	&  305466.02    &   {\bf 304613.18}\\	
&& {\bf LCcw}  	   & 316998.91 &{\bf 303623.67}  &   303759.81  & 303910.04		& 304025.80 \\		
\hline
\hline 
\end{tabular}
\caption{\footnotesize {\bf Model selection with BIC} computed for each model at each data generating model - LCreg, LCdist and LCcw models - for $S=1,\dots,5$. Data generating model and minimum BIC value, for each model at each scenario, in bold.  \label{table:BICcomp}}
\end{table}
 \FloatBarrier 
\section{The different modeling approaches in details}\label{sec:models}
\subsection{The latent class cluster-weighted model}
Let $\Y = (Y_1,\dots,Y_J)'$ be the vector of the full response pattern and $\y$ its realization. Let us assume also one continuous external variable $Z$ is available, and we denote as $z$ its realization. Let us denote as $X$ the categorical latent variable, with latent classes $s=1,\dots,S.$ A general form of association between $\Y,$ $X$ and $Z,$ involves modeling the following joint probability
\begin{equation}\label{eq:fulljoint}
P(Z=z,X=s,\Y=\y) = P(Z=z,X=s) P(\Y=\y | Z=z,X=s),
\end{equation} 

where the common assumption in LCA of $\Y$ and $Z$ being conditionally independent given the latent process is relaxed. From Equation \eqref{eq:fulljoint}, several submodels can be specified (covariate model, distal outcome model, etc).  If substantive theoretical arguments postulate the latent variable to be a predictor of the external variable $Z,$ the latent class cluster-weighted model specifies the probability of observing a response pattern $\y$ as
\begin{equation}\label{eq:cwmod}
P(\Y = \y , Z=z) = \sum_{s=1}^S \underbrace{P(X=s)}_{\text{a}} \underbrace{P(Z=z | X=s)}_{\text{b}} \underbrace{P(\Y=\y | Z=z,X=s)}_{\text{c}},
\end{equation}
which is defined by three components: the structural component (a), which describes
the latent class variable, a measurement
component (b), connecting the latent class to the observed responses with a direct effect of $Z$, and the external variable model (c), which models the latent class specific distribution of $Z$.
Under the assumption of local independence of response variables given the class membership and $Z$,
the conditional distribution of the responses can be
written as
\begin{equation}\label{eq:locind}
 P(\Y=\y | Z=z,X=s) = \prod_{j=1}^J P(Y_j=y_j | Z=z,X=s).
 \end{equation}

For estimating the model in Equation \eqref{eq:cwmod}, we assume each $Y_j$ to be conditionally Bernoulli distributed, with success probability $\pi_{sj},$ and parametrize the conditional response probabilities through the following log-odds
\begin{equation}\label{eq:condrprobcwm}
\log \left( \frac{\pi_{js}}{1-\pi_{js}} \right) = \beta_{0,js} + Z\beta_{js},
\end{equation}
whereby $Z$ is assumed to be conditionally Gaussian with mean $\mu_s$ and variance $\sigma^2_s,$  for  $1 \leq   s \leq S.$ 

The model of Equation \eqref{eq:cwmod} can be used to assign observations to clusters based on the posterior membership probabilities
\begin{equation}\label{eq:postcwm}
P(\text{X} = s | \Y = \mathbf{y},Z=z) = \frac{P(X=s) P(Z=z | X=s) P(\Y=\y | Z=z, X=s)}{P(\Y = \y , Z=z)},
\end{equation}
according to, for instance, modal or proportional assignment rules. 

The latent class unconditional probabilities can as well be parametrized using logistic regressions. We opt for the following parametrization
\begin{equation}\label{eq:mixprop}
\log \left( \frac{P(X=s)}{P(X=1)} \right) = \theta_{s},
\end{equation}
for $1 <   s \leq S,$ where we take the first category as reference, and we set to zero the related parameter. The total number of free parameters to be estimated is therefore $2(J \times S)$ for the measurement model, $2S$ for the external variable model, and $S-1$ for the structural model.

Notice that, by setting the $\beta_{js}$'s of Equation \eqref{eq:condrprobcwm} to zero, the LC cluster-weighted model reduces to a standard LC with distal outcome model. By contrast, given that the external variable component is completely missing, the LC regression is not formally nested in the LC cluster-weighted model, although it can be thought of as a sub-model in which the conditional distribution of $\mathbf{Y} | Z$ is modeled, and $Z$ is taken as fixed-value rather than a random variable to be modeled as well.    

\subsection{The LC with distal outcome model}
It is common, in LCA, to consider a less general version of the joint distribution of Equation \eqref{eq:fulljoint}, by assuming the responses and $Z$ to be conditionally independent given the latent process. If, again, the latent class variable is taken to be a predictor of the external variable $Z,$ this yields the following latent class with distal outcome model
\begin{equation}\label{eq:LCdisteq}
P(\Y = \y , Z=z) = \sum_{s=1}^S P(X=s) P(Z=z | X=s) P(\Y=\y | X=s).
\end{equation}
Under the local independence assumption of the items given the latent class variable, the response conditional probabilities can be written, similarly to Equation \eqref{eq:locind}, as
\begin{equation}\label{eq:locindDist}
 P(\Y=\y | X=s) = \prod_{j=1}^J P(Y_j=y_j | X=s),
 \end{equation}
and parametrized through the following log-odds
\begin{equation}\label{eq:condrprobDist}
\log \left( \frac{\pi_{js}}{1-\pi_{js}} \right) = \beta_{0,js}.
\end{equation}

The model of Equation \eqref{eq:LCdisteq} can be used to cluster observations, according to modal or proportional assignment rules, based on the following posterior membership probabilities
\begin{equation}\label{eq:postdist}
P(\text{X} = s | \Y = \mathbf{y},Z=z) = \frac{P(X=s) P(Z=z | X=s) P(\Y=\y | X=s)}{P(\Y = \y , Z=z)}.
\end{equation}

The external variable $Z$ is assumed, conditional to the latent class, to be Gaussian with mean $\mu_s$ and variance $\sigma^2_s,$ for $1 \leq s \leq S,$ whereby the latent class unconditional probabilities are parametrized as in Equation \eqref{eq:mixprop}. This yields $J \times S + 2S + S-1$ free parameters to be estimated. The only difference with the model of Equation \eqref{eq:cwmod} is that $\mathbf{Y},$ in the measurement component, is conditional only on $X$, not on $Z$. That is, $\mathbf{Y}$ is assumed to be independent of $Z$ given $X$, which is a standard, and rather very strong, assumption of LCA.

\subsection{The LC regression model}

Rather than modeling the joint distribution $P(Z,X,\Y)$ of Equation \eqref{eq:fulljoint}, the latent class regression models the conditional distribution of $\Y$ given $Z$ and the latent class variable, specifying the following model for $\Y$:
\begin{equation}\label{eq:LCregeq}
P(\Y = \y | Z=z) = \sum_{s=1}^S P(X=s)  P(\Y=\y | Z=z,X=s).
\end{equation}

In this case, the conditional response probabilities depend on the external variable $Z$ and, under local independence of the responses given the latent variable,  the measurement model can be written as in Equation \eqref{eq:locind}, and parametrized as in Equation \eqref{eq:condrprobcwm}. The posterior membership probabilities, computed based on the model in Equation \eqref{eq:LCregeq}, are as follows
\begin{equation}\label{eq:postLCreg}
P(\text{X} = s | \Y = \mathbf{y},Z=z) = \frac{P(X=s) P(Z=z | X=s) P(\Y=\y | X=s)}{P(\Y = \y , Z=z)}.
\end{equation}

With latent class unconditional probabilities parametrized as in Equation \eqref{eq:mixprop}, the total number of free parameters to be estimated is $2(J \times S) + S-1.$
\FloatBarrier

\begin{table}[!h]
\centering  
\begin{tabular}{lcccc}
\hline \hline
									&	 & Dir. Eff. & $Z$ modeled & \#par \\
\hline
Latent Class regression 			&	& $\checkmark$	& $\times$ & $2(J \times S) + (S-1)$\\
Latent Class with distal outcome 	&	& $\times$	& $\checkmark$ & $(J \times S) + 2S + (S-1)$\\
Latent Class cluster-weighted 		&	& $\checkmark$ & $\checkmark$ & $2(J \times S) + 2S + (S-1)$ \\
\hline \hline
\end{tabular}
\caption{Summary of different modeling assumptions and number of free parameters to be estimated.\label{table:modsumassumptions}}
\end{table} 
\FloatBarrier

Table \ref{table:modsumassumptions} summarizes how $Z$ enters each of the three models, and the total number of free parameters to be estimated. Intuitively, this shows that the first two models can be seen as special cases of the third model, which therefore models the relationship between the 3 sets of variables in the most exhaustive manner.
\section{A latent class model of households' assets ownership to predict wealth}\label{sec:empirical}

Household wealth cannot be directly observed. Nonetheless, measuring it is a crucial issue for any policy maker. Notably, survey measures of income - or expenditure, if available - are affected by substantial measurement error and systematic reporting bias \citep{ferguson2003,moore2000}. In addition, if wealth as a measure of permanent income \citep{friedman1957} is of interest, current income, even if measured without error, is likely to be a poor approximation. In more recent surveys (like the Household Finance and Consumption Survey, from the European Central Bank), a measure of net wealth - value of total household assets minus the value of total liabilities - is provided. However, relying on each household's subjective evaluation of the current value of each asset they own, such a measure is prone to considerable measurement error as well. Furthermore, having such a complex measure is complicated to understand for a more general audience, to whom a simpler classification/index would appeal.     

Latent Trait (LTA) and Latent Class Analysis have been used in order to model wealth - or, inversely, deprivation - from observed assets ownership. Whereby in the LTA framework, wealth is modeled as a continuous trait \citep{szeles2013,vandemoortele2014}, LCA was used \citep{moisio2004,perez2005} based on the idea that wealth (poverty) can be seen as a multidimensional latent construct. Although determining which ownership indicators to use can be a problem, arguably they all attempt to identify subgroups in the population based on the same multidimensional phenomenon \citep{moisio2004}: different dimensions of wealth are measured by different (sets of) indicators. In addition, within an LCA framework, response probabilities can be used to evaluate \emph{ex post} each indicator's validity of measuring the (latent variable) wealth. 

We analyze data from the first wave of the Household Finance and Consumption Survey, conducted by the European Central Bank. We focus on a sample of Italian households, for which we have information on real and financial assets, liabilities, different income measures, consumption expenditures, and a measure of total wealth in euro. The latter is defined as total household assets value, excluding public and occupational pension wealth, minus total outstanding household's liabilities \citep{ecb2013}. The value of each asset is provided by asking to the interviewees how much they think each asset is worth. For instance, related to the item ``owning any car" (HB4300, Table \ref{table:varlist}), the interviewer asks ``For the cars that you/your household own, if you sold them now, about how much do you think you could get?" (HFCS Core Variables Catalogue, 2013). In fact, different households might have very different (and possibly wrong) perceptions about the value of the assets they own. This is why we cannot rely only on DN3001 as a measure of households wealth, but rather use a model known for its strengths in correcting for measurement error using multiple indicators. 

We selected a set of 10 items related to a financial type of wealth, and we included also 4 items related to a broader type of wealth, for a total of 14 items. The variable concerning household residence tenure status (HB0300) was recoded to have ``entirely owned main residence" or ``partially owned main residence" merged into one category. The resulting variable on tenure status (hometen) has 3 categories, and enters all models with dummy coding. 

We restrict ourselves to analyzing only households having positive wealth. Doing so allows us to set up a model for log-wealth rather than for wealth, as is commonly done in the economic literature studying elasticities (see, for instance, \citealp{charles2003}). Imposing this restriction leads us to drop only 188 sample units (about 2\% of the total sample).   

The LCreg, the LCdist and the LCcw models are estimated using Latent GOLD 5.1 \citep{latentG5.1}. Sample syntaxes for each model are reported in the Appendix. The model comparison is done with the purpose of investigating how cluster membership is able to predict classes of (log) wealth. The cluster-weighted modeling approach does so by relaxing the conditional independence assumption and allowing for possible direct effects of log-wealth on the indicators. 

\begin{table}[!h]
\centering  
\begin{tabular}{lll}
\hline \hline
Name & Description & Type\\
\hline
HB0300 & Household main residence - Tenure status	& Nominal \\
	   & (1 if entirely owned)						&			 \\
	   & (2 if partially owned)						&			 \\
	   & (3 if rented)								&			 \\
	   & (4 if for free use)						&			 \\
hometen & Household main residence - Tenure status	& Nominal \\
	   & (1 if entirely or partially owned)			&			 \\
	   & (2 if rented)								&			 \\
	   & (3 if for free use)						&			 \\ 	   
HB2400 & Household owns other properties   			& Dichotomous \\
HB4300 & Household owns any car			   			& Dichotomous \\
HB4700 & Ownership of other valuables      			& Dichotomous \\
HC0200 & Household has a credit line or overdraft	& Dichotomous \\
HC0300 & Household has a credit card				& Dichotomous \\
HC0400 & Household has a non collaterized loan		& Dichotomous \\
HD0100 & Household has any investment in business	& Dichotomous \\
HD1100 & Household owns a sight account				& Dichotomous \\
HD1200 & Household owns a savings account			& Dichotomous \\
HD1300 & Household has any investment in mutual fund& Dichotomous \\
HD1400 & Household owns bonds						& Dichotomous \\
HD1500 & Household owns managed accounts			& Dichotomous \\
DN3001 & Net Wealth									& Continuous \\
	   
\hline
\hline
\end{tabular}
\caption{\footnotesize Variables list, with description and type. hometen obtained by recoding HB0300 into 3 categories, where ``partially owned" and ``entirely owned" are merged into one. DN3001 is obtained as total household assets, excluding public and occupational pension wealth, minus total outstanding household's liabilities. We focus on a sample of observations with positive net wealth, and work with its logarithmic transformation. Further details on variables definitions are available from the ECB website.\label{table:varlist}}
\end{table}

In Table \ref{table:modselval} we display number of components, BIC, number of parameters, expected classification error, and entropy-based $R^2$ for each of the three models, from 1 to 5 components. Expected classification error is a standard output in Latent GOLD, and is obtained by cross-tabulating the modal classes (as estimated by the modal assignment rule)  with probabilistic classes (as estimated by the class proportions), then taking the ratio between the number of missclassified observations and the total number of units.  

\begin{table}[!h]
\centering  
\begin{tabular}{lccccc}
\hline \hline
		&	$S$	&	BIC	&	\#par	&	Class. Err.	&	Entr. $R^2$	\\
		\hline
\multirow{5}{*}{LCreg}	&	1	&	84730.90	&	27	&	-	&	-	\\
	&	2	&	80709.36	&	55	&	0.12	&	0.59	\\
	&	3	&	79157.86	&	83	&	0.18	&	0.58	\\
	&	4	&	78647.04	&	111	&	0.24	&	0.57	\\
	&	5	&	78680.09	&	139	&	0.25	&	0.57	\\
	\cmidrule{2-6}
\multirow{5}{*}{LCdist}		&	1	&	125830.86	&	16	&	-	&	-	\\
	&	2	&	114860.57	&	33	&	0.06	&	0.77	\\
	&	3	&	109951.62	&	50	&	0.08	&	0.83	\\
	&	4	&	108642.01	&	67	&	0.11	&	0.81	\\
	&	5	&	107201.39	&	84	&	0.14	&	0.79	\\
	\cmidrule{2-6}
\multirow{5}{*}{LCcw}	&	1	&	115507.90	&	29	&	-	&	-	\\
	&	2	&	108412.34	&	59	&	0.01	&	0.95	\\
	&	3	&	106477.23	&	89	&	0.09	&	0.80	\\
	&	4	&	105996.46	&	119	&	0.17	&	0.72	\\
	&	5	&	105731.14	&	149	&	0.19	&	0.73	\\

\hline
\hline
\end{tabular}
\caption{\footnotesize Number of components (Ncomp), BIC, number of parameters (\#par), expected classification error (Class. Err.), and entropy-based (Entr.) $R^2$ for each of the three models, from 1 to 5 components.\label{table:modselval}}
\end{table}

Based on BIC only, both LCdist and LCcw models would select the highest possible number of components ($S=5$), whereas LCreg would select four components. In fact, consistently with previous literature on LCA for measuring wealth \citep{moisio2004,perez2005}, both with four and five components, latent classes would be hardly interpretable. $S=2$ seems to be the best trade-off solution according to BIC and the two fit criteria for the LCreg and LCcw models. LCdist seems to favor $S=3$. We select 2 classes for LCreg and LCcw, and 3 classes for LCdist. We also show, for comparability reasons, results for LCdist with $S=2$. Interestingly however, BIC values of LCcw are always smaller than those of LCdist, indicating a better overall fit to the data.

\begin{figure}[!h]
\captionsetup{font={footnotesize}}
\begin{center}
\subfloat[LCreg, $S=2$]{\label{fig:LCreg2profile}\includegraphics[width=0.45\textwidth]{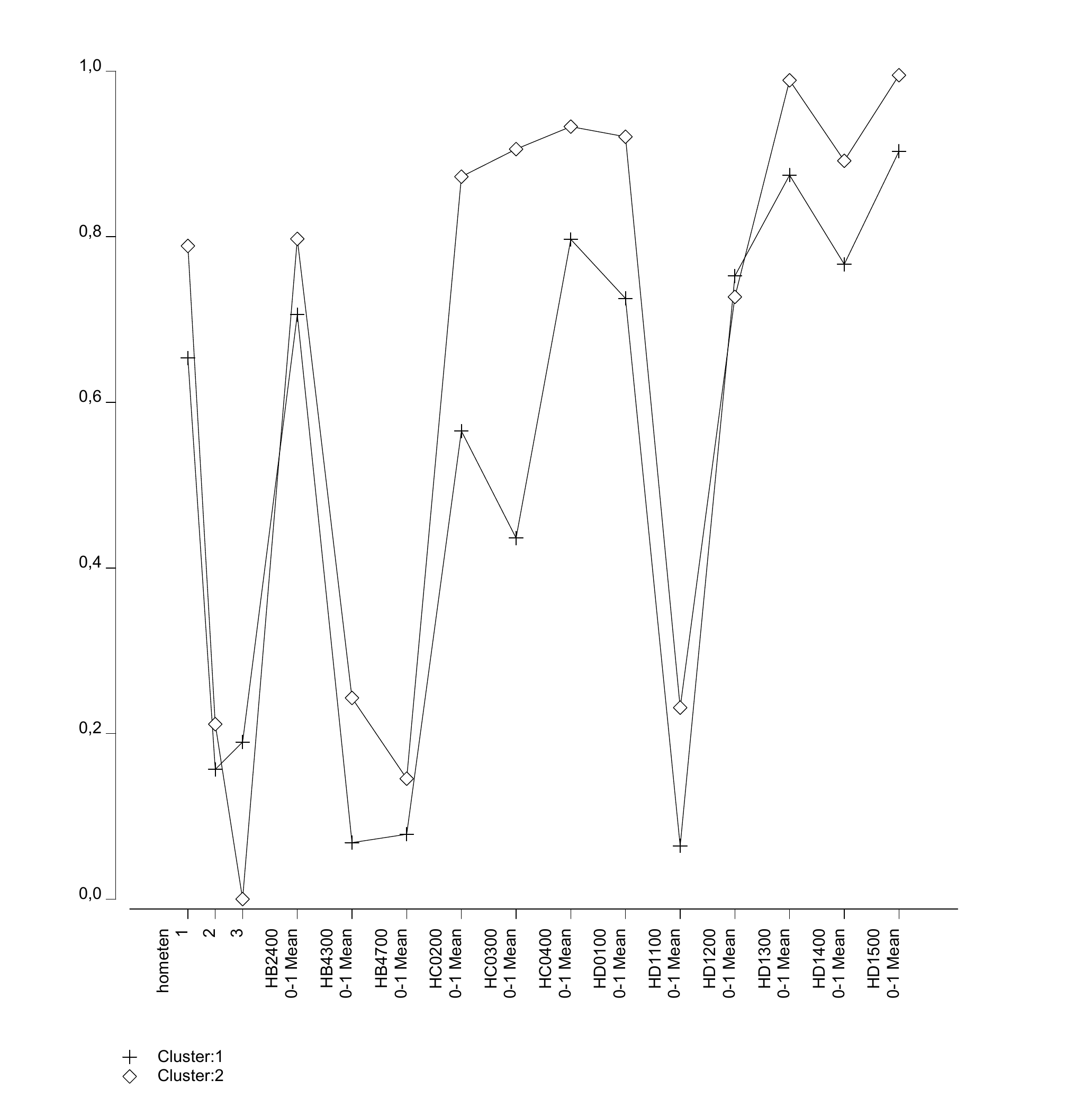}}\qquad
\subfloat[LCcw, $S=2$]{\label{fig:LCcw2profile}\includegraphics[width=0.45\textwidth]{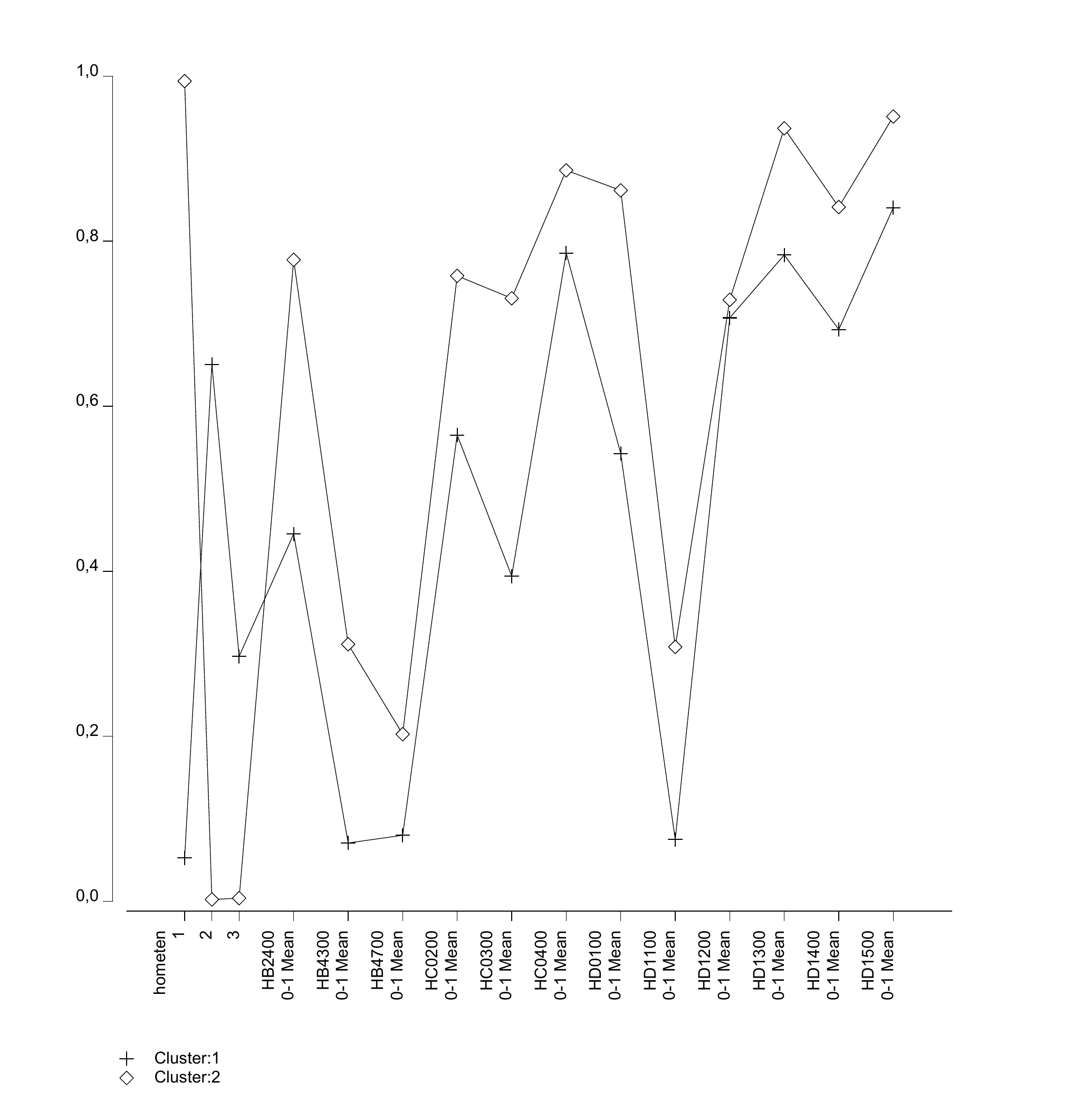}}\\
\subfloat[LCdist, $S=2$]{\label{fig:LCdist2profile}\includegraphics[width=0.45\textwidth]{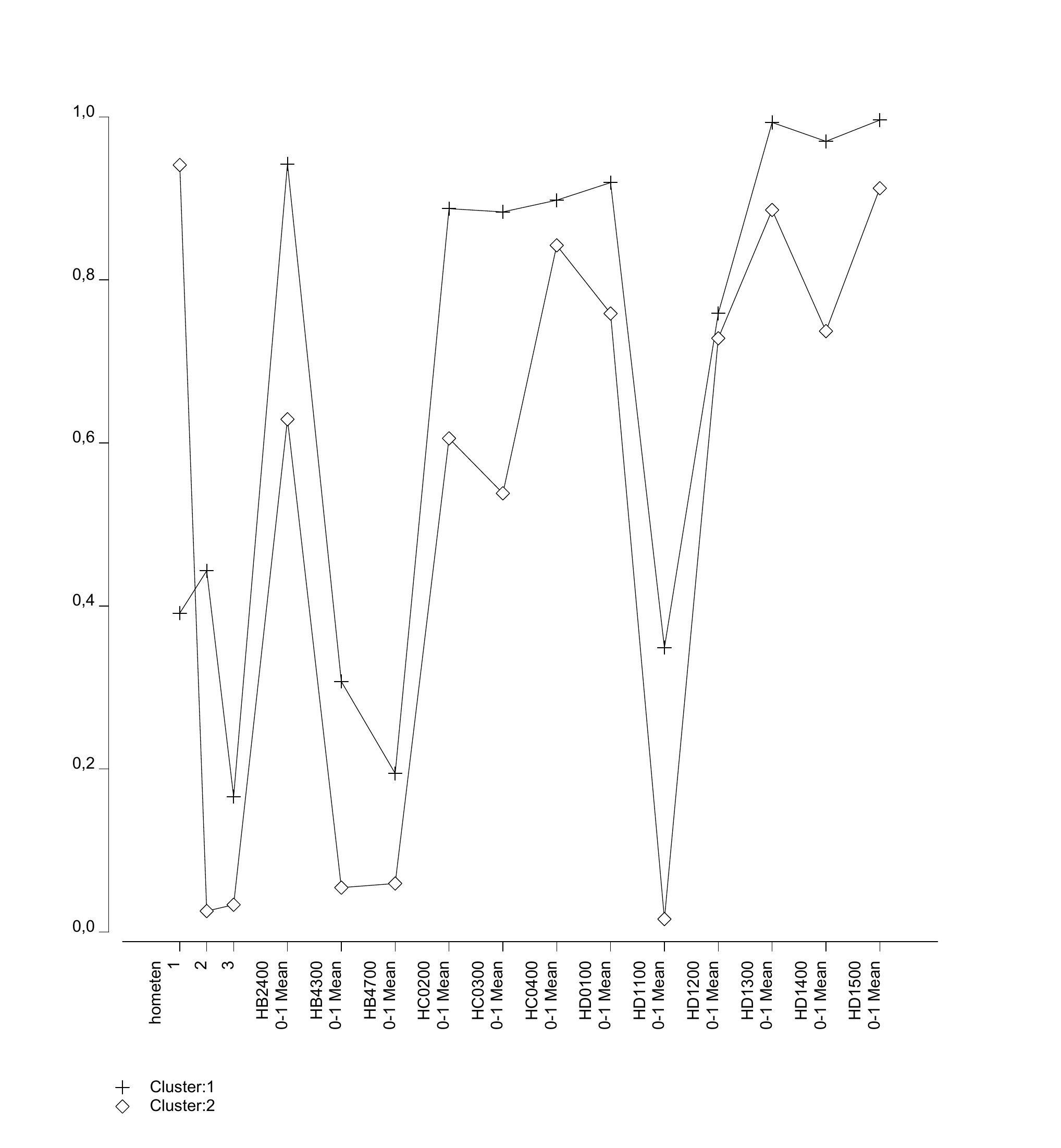}}\qquad 
\subfloat[LCdist, $S=3$]{\label{fig:LCdist3profile}\includegraphics[width=0.45\textwidth]{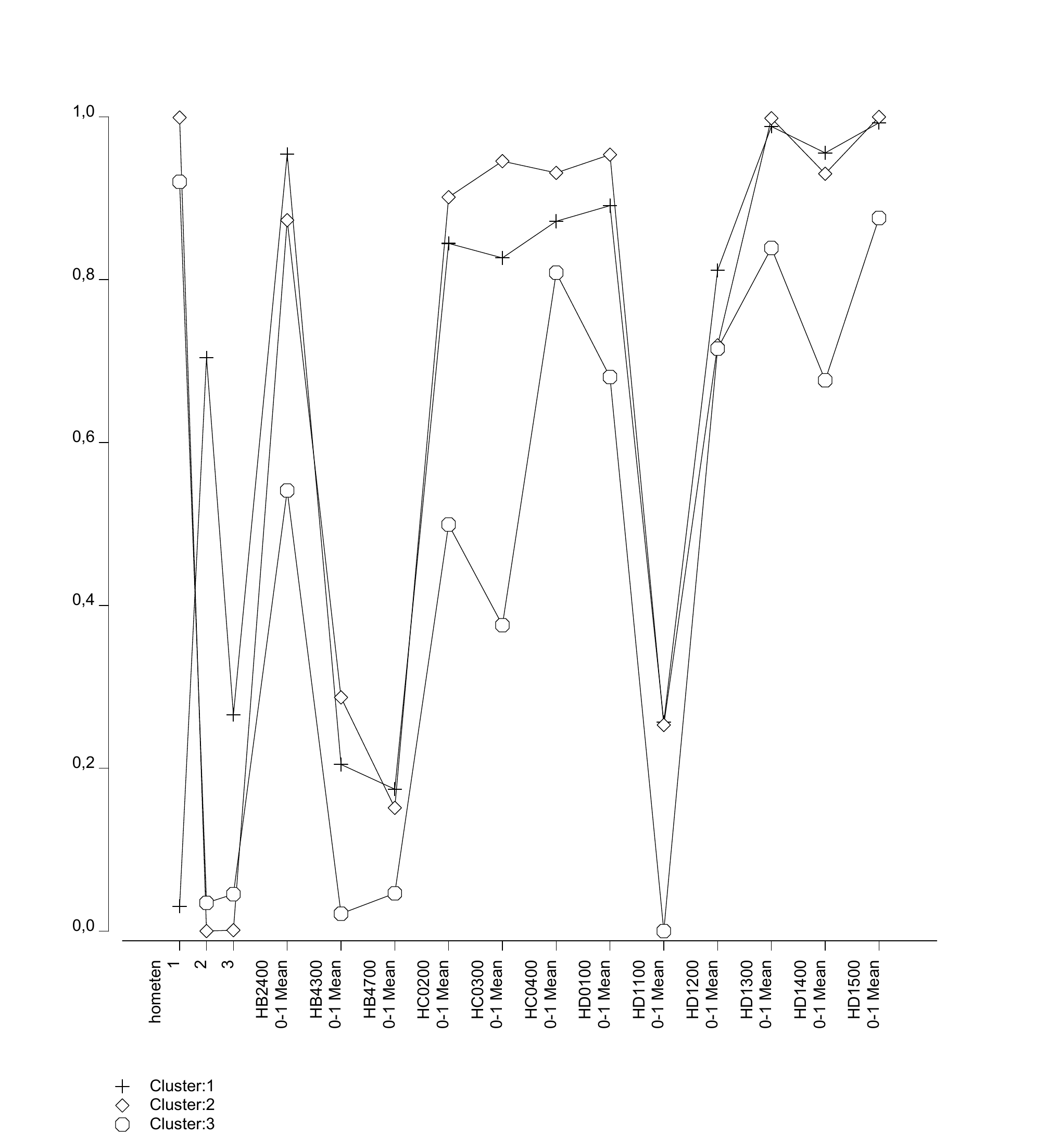}}

\caption{\footnotesize{Probability profile plot for LCreg $(S=2)$, LCdist $(S=2$ and $S=3$) and LCcw $(S=2)$. For the variable  ``hometen", the levels indicate the four category ownership probabilities within each wealth class. Similarly, levels refer to average item ownership probability in wealth classes for the remaining (dichotomous) items.}}
\label{fig:profiles}
\end{center}
\end{figure}

We report (Figure \ref{fig:profiles}) the probability profile plot for LCreg with $S=2$ (Figure \ref{fig:LCreg2profile}), LCdist for $S=2$ and $S=3$(Figures \ref{fig:LCdist2profile} and \ref{fig:LCdist3profile} respectively), and LCcw for$S=2$ (Figure \ref{fig:LCcw2profile}).

We observe that both the LCreg and the LCcw models, contrary to the two-class LCdist model, predict a wealthier class, with higher ownership probabilities for all items, and a higher probability of owning (partially or entirely) the household main residence - relative to ``free use" and ``rent" categories. In the three-class LCdist model, whose first and second clusters seem to arise from a split of the first cluster of the two-class model, the first wealth class shows higher ownership probabilities for all items, and a higher probability of owning (partially or entirely) the household main residence, as in the two-class LCreg and LCcw models. One possible explanation for this is that the LCdist model needs one additional class to predict a class profile reasonably corresponding to the highest wealth level.
Interestingly, profiles of all three models ($S = 2$) are comparable, though each showing specific features. LCreg classes have a similar composition in terms of profiles. Both LCdist ($S = 2$ and $S = 3$) and LCcw deliver better separated classes. However, wealthier households are predicted by LCcw to own more assets, whereas less wealthy household have a relatively larger probability, for instance, to have a rented (or for free use) residence.

Table \ref{table:ARIempirical} reports the ARI for pairwise comparisons of the three models. LCdist, with both $S=2$ and $S=3$, and LCcw deliver clusterings with moderate agreement. Figure \ref{fig:logwealth} shows how the class composition of LCdist (with $S=2$ and $S=3$) and LCcw predicts classes in log-wealth.
 
\begin{table}[!h]
\centering  
\begin{tabular}{lcccc}
\hline \hline
		& LCreg	   &	LCdist   & LCdist &    LCcwm	 	\\
		&			&   $(S = 2)$ & $(S = 3)$ & \\
		     
\hline

LCreg  			&	 1		   &	  	&  &\\	
LCdist $(S = 2)$&	 0.0869	   &	  1 &    &  \\	
LCdist $(S = 3)$&	 0.3369	   &	0.4446 &  1  & 	\\	
LCcw  			&	 0.0661    &	  0.4676    &  0.4371  & 1	\\	
\hline
\hline 
\end{tabular}
\caption{\footnotesize {\bf Adjusted Rand indexes} comparing clustering in classes of wealth of LCreg, LCdist and LCcw models.\label{table:ARIempirical}}
\end{table}

\begin{figure}[!h]
\captionsetup{font={footnotesize}}
\begin{center}
\subfloat[LCdist, $S=2$]{\label{fig:LCdist2logwealth}\includegraphics[width=0.45\textwidth]{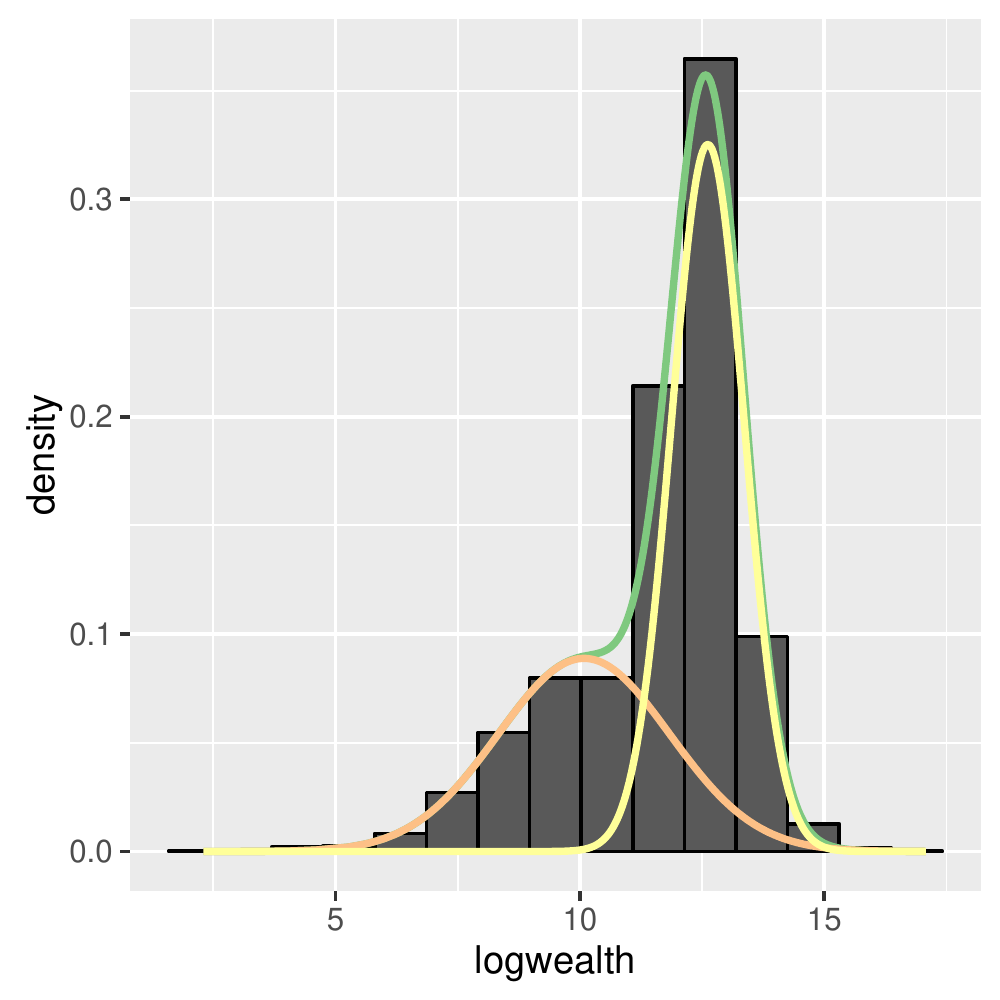}}\qquad
\subfloat[LCdist, $S=3$]{\label{fig:LCdist3logwealth}\includegraphics[width=0.45\textwidth]{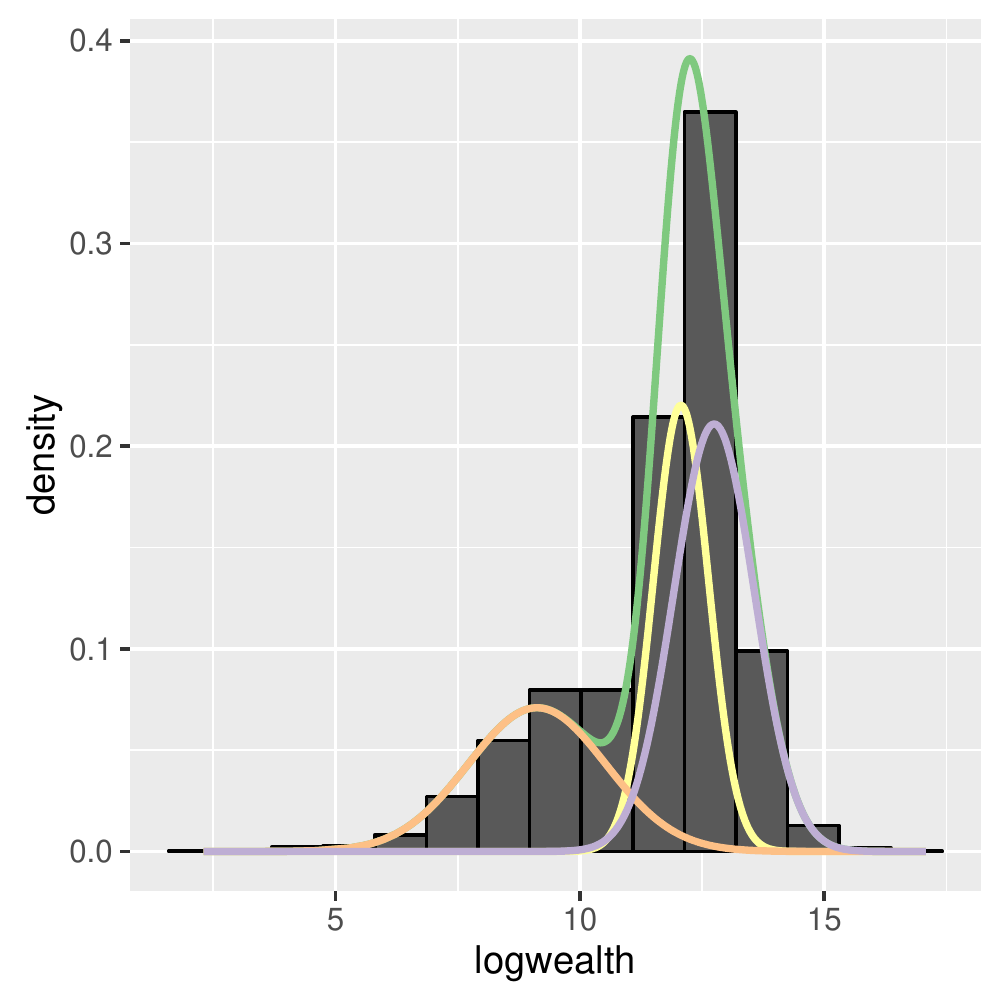}}\\
\subfloat[LCcw]{\label{fig:LCcwlogwealth}\includegraphics[width=0.45\textwidth]{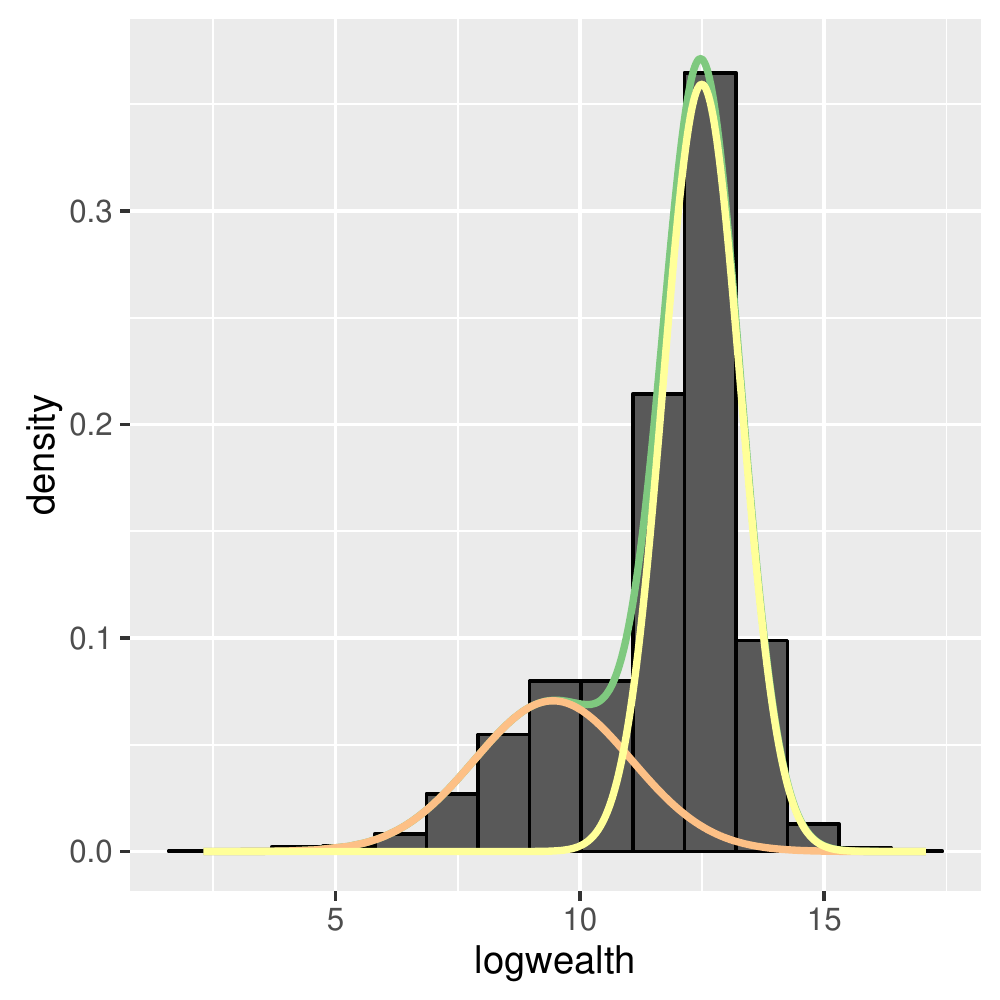}} \qquad
\subfloat[Class proportions]{\label{fig:class_prop_bar}\includegraphics[width=0.45\textwidth]{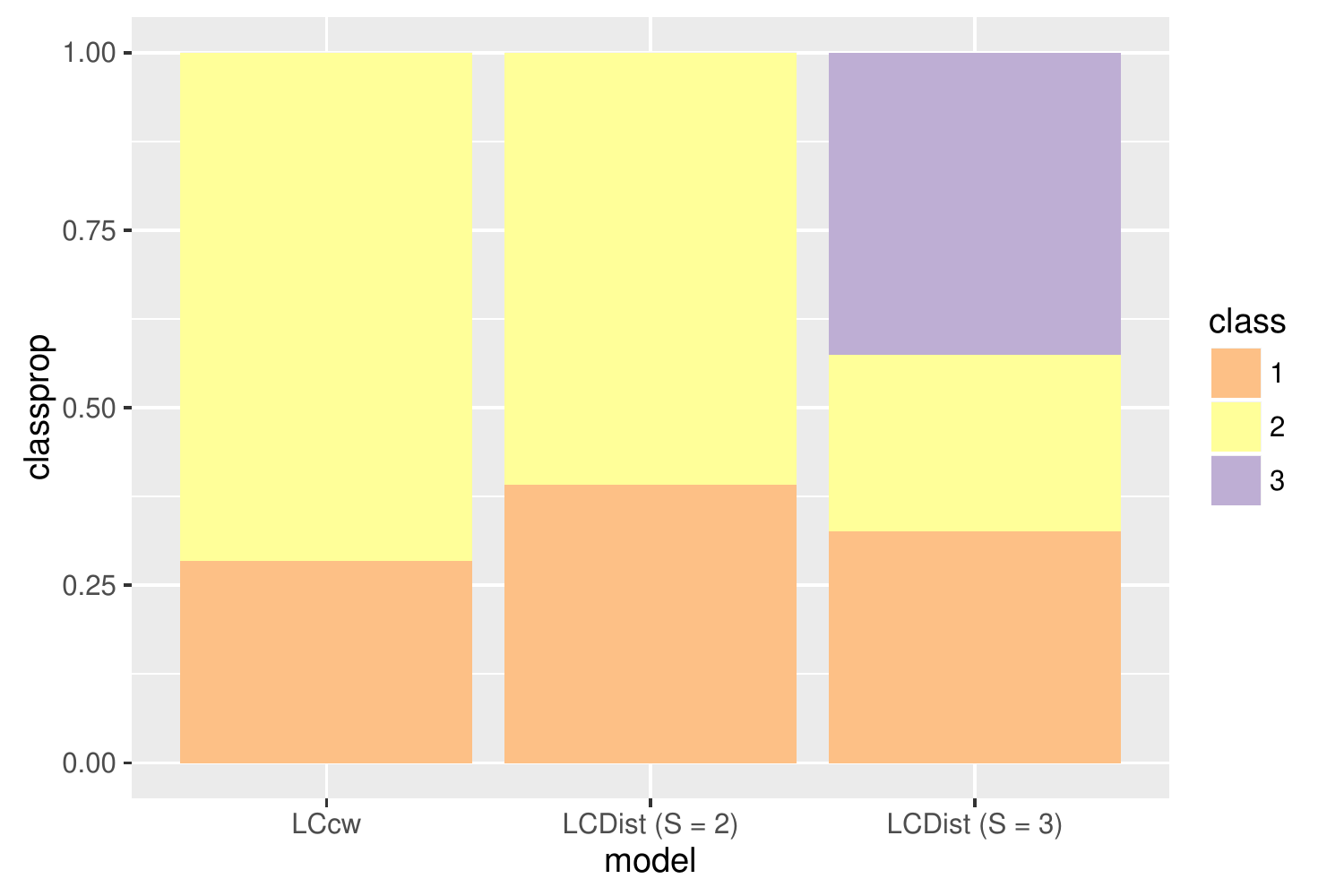}}
\label{fig:logwealth}
\caption{\footnotesize{Density plots of log-wealth as predicted by LCdist $(S=2$, a, and $S=3$,b) and LCcw (c). Mixture density in green, component marginal densities in other colors. Class proportions in (d).}}
\label{fig:logwealth}
\end{center}
\end{figure}

The first class in LCdist with $S = 2$ (Figure \ref{fig:LCdist2logwealth}) seems to capture observations with lower log-wealth, although its relatively fat tails allocate non-zero density also to the wealthiest observations. With $S = 3$ (Figure \ref{fig:LCdist3logwealth}), the second class is split into two (see also Figure \ref{fig:class_prop_bar}), with the new third class made up of households in the right tale of log-wealth distribution. Consistently with findings in the probability profiles, classes in LCcw are able to discriminate households better in terms of prediction of the (log) wealth distribution (Figure \ref{fig:LCcwlogwealth}). 

To gain further insights on the estimated wealth distribution, we report (Table \ref{table:logwealthdist}) estimated means and variances of log-wealth, for LCdist with $S=2$ and LCcw. For comparability, we choose not to report mean and variance values for LCdist with $S = 3$. 

\begin{table}[!h]
\centering  
\begin{tabular}{lcccccc}
\hline \hline
& \multicolumn{2}{c}{Means} & Wald(=) $p$ &  \multicolumn{2}{c}{Variances} &  Wald(=) $p$ \\
\hline
LCdist&	 10.0835***&	  12.6173*** &    0.0000 	&   3.0910 &    0.5573 & 0.0000 \\	
&	(0.0451)		&	(0.0129)	&	  		&   (0.0843)   & (0.0137)     &     \\	
&					&				&			&   		   & 		     &     \\	
LCcw  &  9.4489***	&	12.4928***  &	0.0000  &   2.5960   & 0.6304       &  0.0380  \\	
&			(0.0391)  &     (0.0112)&			&    (0.0865)  &     (0.0131)     &     \\	
\hline
\hline 
\end{tabular}
\caption{\footnotesize Estimated means (*** $p$-value$<$0.01, ** $p$-value$<$0.05, * $p$-value$<$0.1) and variances of log-wealth, and $p$-values from Wald test of equality of component means and variances for the LCdist $(S = 2)$ and LCcw models. Standard errors in parentheses. \label{table:logwealthdist}}
\end{table}

Inference on the estimated parameters of the log-wealth ($p$-values of all tests are below 0.01) validates the model assumption of clustered distribution, with heteroscedastic components, of log-wealth in both LCdist and LCcw. The mean in the first class of wealth is larger for LCdist than for LCcw, along with the related variance. As we observed above, this first class in LCdist absorbs also households in the right tale of the distribution. This complicates the interpretation of the wealth classes compared to LCcw. That is, LCcw, with a first class with smaller mean and variance, allows for a more natural substantive interpretation of the two classes as wealthier versus less wealthy households. This is also consistent with findings of the related literature on poverty (deprivation) \citep{moisio2004,perez2005}.

Finally, Table \ref{table:directeff} demonstrates that LCcw provides an easy way to test whether there are significant direct effects of log-wealth on the response variables. Interestingly, inference on such effects points out that there are variables for which we have no significant direct effects of log-wealth (hometen), as well as indicating that some effects are the same across latent classes (HB2400, HD1300, HD1400 and HD1500). In the logic of an applied researcher, this suggests intermediate and more parsimonious modeling options, where effects on some variables can be constrained to be zero, or to be the same across classes, which can easily be discovered with a cluster-weighted modeling approach.

\begin{table}[!h]
\centering  
\resizebox{0.68\textheight}{!}{\begin{tabular}{lcccc}
\hline \hline
Log-wealth on & \multicolumn{2}{c}{Coefficients} & Wald(0) $p$ &  Wald(=) $p$ \\
			& Class 1 & Class 2 				&				&				\\
\hline
hometen	&	0.0184		&	0.1532		&	0.6800	&	0.4800	\\
(Main res. tenure stat)	&	(0.0392)		&	(0.1930)		&		&		\\
	&			&			&		&		\\

HB2400	&	-1.6242***	&	-1.5421***	&	0.0000	&	0.4600	\\
(HH owns other properties)	&	(0.0957)		&	(0.0564)		&		&		\\
	&			&			&		&		\\

HB4300	&	-0.5784***	&	-1.1729***	&	0.0000	&	0.0000	\\
(HH owns any car)	&	(0.0426)		&	(0.0626)		&		&		\\
	&			&			&		&		\\

HB4700	&	-0.3866***	&	-0.7445***	&	0.0000	&	0.0000	\\
(HH owns other valuables)	&	(0.0396)		&	(0.0639)		&		&		\\
	&			&			&		&		\\

HC0200	&	-0.5217***	&	-0.7899***	&	0.0000	&	0.0000	\\
(HH has a credit line)	&	(0.0431)		&	(0.0433)		&		&		\\
	&			&			&		&		\\

HC0300	&	-0.8741***	&	-1.1398***	&	0.0000	&	0.0001	\\
(HH has a credit card)	&	(0.0517)		&	(0.0480)		&		&		\\
	&			&			&		&		\\

HC0400	&	-0.2026***	&	-0.2205***	&	0.0000	&	0.7800	\\
(HH has a non-coll. loan)	&	(0.0415)		&	(0.0515)		&		&		\\
	&			&			&		&		\\

HD0100	&	-0.7896***	&	-1.0633***	&	0.0000	&	0.0002	\\
(HH has any invest. in business)	&	(0.0536)		&	(0.0527)		&		&		\\
	&			&			&		&		\\

HD1100	&	-0.8060***	&	-1.6133***	&	0.0000	&	0.0000	\\
(HH owns a sight account)	&	(0.0463)		&	(0.0768)		&		&		\\
	&			&			&		&		\\

HD1200	&	-0.2216***	&	-0.0630*	&	0.0000	&	0.0025	\\
(HH owns a savings account)	&	(0.0364)		&	(0.0381)		&		&		\\
	&			&			&		&		\\

HD1300	&	-1.0827***	&	-1.1224***	&	0.0000	&	0.7500	\\
(HH has invest. in mutual funds)	&	(0.1061)		&	(0.0636)		&		&		\\
	&			&			&		&		\\

HD1400	&	-0.9244***	&	-0.9835***	&	0.0000	&	0.5100	\\
(HH owns bonds)	&	(0.0766)		&	(0.0478)		&		&		\\
	&			&			&		&		\\

HD1500	&	-1.0942***	&	-1.1664***	&	0.0000	&	0.6100	\\
(HH owns managed accounts)	&	(0.1236)		&	(0.0688)		&		&		\\
 \\	
\hline
\hline 
\end{tabular}}
\caption{\footnotesize Estimated direct effects of log-wealth on each variable per latent class (*** $p$-value$<$0.01, ** $p$-value$<$0.05, * $p$-value$<$0.1), $p$-values from Wald test of joint equality of each variable's direct effect to zero - Wald(0) - and from Wald test of equality of effects across latent classes - Wald(=) - for the LCcw model. Short description below variable names. HH stands for Household. Standard errors in parentheses.  \label{table:directeff}}
\end{table}
\FloatBarrier
\section{Conclusion}\label{sec:concl}

In this paper we have brought modeling ideas from the regression mixture literature into latent class analysis. Our focus has been to motivate the use of the cluster-weighted modeling approach as a general specification for the joint relationship of the response variables, the external variable, and the latent variable. Three population studies have been used to illustrate this idea, and the actual advantage of the proposed approach was showed through an application on Italian household asset ownership data.

The cluster-weighted modeling approach, contrary to the simpler distal outcome model, was able to predict, based on the joint relationship of ownership indicators, the underlying true wealth, and wealth as measured by the survey, two interpretable classes of wealth with class ownership profiles coinciding respectively with less wealthy (first class) and wealthier (second class) households on the measured log-wealth distribution.  

In the applied researcher perspective, the proposed approach has several advantages. First, it allows to deal with direct effects if this are of substantive interest, as well as if those represent a source of noise to be handled. That is, if direct effects are present, our approach, contrary to the distal outcome model, yields unbiased estimates of the distal outcome cluster specific means and variances. Second, it guarantees a safe and flexible option, in which one starts from the most general model. Then, proceeding backwards, the user can test the model assumptions of both the distal outcome and the latent class regression models.

The approach we suggest has some limitations as well. First, it can be unstable if some of the observed response patterns are lacking, or are simply too small in number to estimate the model parameters (see for instance, for standard sufficient conditions for identifiability of LCA, \citealp{bandeenroche:97}). Especially in an exploratory stage of the analysis, in such cases simpler models can be more attractive. Second, depending on the goal of the analysis, the cluster-weighted modeling approach can generate a final output which might be harder to interpret than that of simpler models. In other words, the final model interpretability relies on the final goal of the analysis, which must be clear in mind in this as well as in any other modeling approach.

\bibliography{azr}
\bibliographystyle{apacite}

\appendix
\section{Latent GOLD syntax for the three models}
\subsection{LCreg model syntax}
\small{options

   maxthreads=4;
   
   algorithm 
   
      tolerance=1e-008 emtolerance=0.01 emiterations=500 nriterations=100 ;
      
   startvalues
   
      seed=0 sets=50 tolerance=1e-005 iterations=50;
      
   bayes
   
      categorical=1 variances=1 latent=1 poisson=1;
      
   montecarlo
   
      seed=0 sets=0 replicates=500 tolerance=1e-008;
      
   quadrature  nodes=10;
   
   missing  excludeall;
   
   output      
   
      parameters=first betaopts=wl standarderrors profile probmeans=posterior
      
      bivariateresiduals estimatedvalues=model;
      
variables

   dependent hometen nominal, HB2400, HB4300, HB4700, HC0200,
   
             HC0300, HC0400, HD0100, 
             
             HD1100, HD1200, HD1300, HD1400, HD1500;
             
  independent logwealth; 
  
   latent
   
      Cluster nominal 2;
      
equations

   Cluster $<-$ 1;
   
hometen $<-$ 1$|$Cluster + logwealth$|$Cluster;

HB2400 $<-$ 1$|$Cluster + logwealth$|$Cluster;

HB4300 $<-$ 1$|$Cluster + logwealth$|$Cluster;

HB4700 $<-$ 1$|$Cluster + logwealth$|$Cluster;

HC0200 $<-$ 1$|$Cluster + logwealth$|$Cluster;

HC0300 $<-$ 1$|$Cluster + logwealth$|$Cluster;

HC0400 $<-$ 1$|$Cluster + logwealth$|$Cluster;

HD0100 $<-$ 1$|$Cluster + logwealth$|$Cluster;

   HD1100 $<-$ 1$|$Cluster + logwealth$|$Cluster;
   
   HD1200 $<-$ 1$|$Cluster + logwealth$|$Cluster;

   HD1300 $<-$ 1$|$Cluster + logwealth$|$Cluster;
   
   HD1400 $<-$ 1$|$Cluster + logwealth$|$Cluster;
   
   HD1500 $<-$ 1$|$Cluster + logwealth$|$Cluster;
   }
\subsection{LCdist model syntax}
\small{options

   maxthreads=4;
   
   algorithm 
   
      tolerance=1e-008 emtolerance=0.01 emiterations=500 nriterations=50 ;
      
   startvalues
   
      seed=0 sets=50 tolerance=1e-005 iterations=50;
      
   bayes
   
      categorical=1 variances=1 latent=1 poisson=1;
      
   montecarlo
   
      seed=0 sets=0 replicates=500 tolerance=1e-008;
      
   quadrature  nodes=10;
   
   missing  excludeall;
   
   output
         
      parameters=first betaopts=wl standarderrors profile probmeans=posterior
      
      bivariateresiduals estimatedvalues=model;
      
variables

   dependent hometen nominal, HB2400, HB4300, HB4700, HC0200,
   
             HC0300, HC0400, HD0100, 
             
             HD1100, HD1200, HD1300, HD1400, HD1500, logwealth continuous; 
             
   latent
   
      Cluster nominal 2;
      
equations

   Cluster $<-$ 1;
   
hometen $<-$ 1$|$Cluster;

HB2400 $<-$ 1$|$Cluster;

HB4300 $<-$ 1$|$Cluster;

HB4700 $<-$ 1$|$Cluster;

HC0200 $<-$ 1$|$Cluster;

HC0300 $<-$ 1$|$Cluster;

HC0400 $<-$ 1$|$Cluster;

HD0100 $<-$ 1$|$Cluster;

   HD1100 $<-$ 1$|$Cluster;
   
   HD1200 $<-$ 1$|$Cluster;
   
      HD1300 $<-$ 1$|$Cluster;
      
   HD1400 $<-$ 1$|$Cluster;
   
   HD1500 $<-$ 1$|$Cluster;
   
logwealth $<-$ 1$|$Cluster;

logwealth$|$Cluster;}
\subsection{LCCw model syntax}
Notice that ``logwealthshadow" is a duplicate of ``logwealth", as Latent GOLD does not allow the same variable to be both dependent and independent variable at the same time.

\small{options

   maxthreads=4;
   
   algorithm 
   
      tolerance=1e-008 emtolerance=0.01 emiterations=500 nriterations=50 ;
      
   startvalues
   
      seed=0 sets=50 tolerance=1e-005 iterations=50;
      
   bayes
   
      categorical=1 variances=1 latent=1 poisson=1;
      
   montecarlo
   
      seed=0 sets=0 replicates=500 tolerance=1e-008;
      
   quadrature  nodes=10;
   
   missing  excludeall;
   
   output
         
      parameters=first betaopts=wl standarderrors profile probmeans=posterior
      
      bivariateresiduals estimatedvalues=model;
      
variables

   dependent hometen nominal, HB2400, HB4300, HB4700, HC0200,
   
             HC0300, HC0400, HD0100, 
             
             HD1100, HD1200, HD1300, HD1400, HD1500, logwealth continuous;
             
  independent logwealthshadow; 
  
   latent
   
      Cluster nominal 2;
      
equations

   Cluster $<-$ 1;
hometen $<-$ 1$|$Cluster + logwealthshadow$|$Cluster;

HB2400 $<-$ 1$|$Cluster + logwealthshadow$|$Cluster;

HB4300 $<-$ 1$|$Cluster + logwealthshadow$|$Cluster;

HB4700 $<-$ 1$|$Cluster + logwealthshadow$|$Cluster;

HC0200 $<-$ 1$|$Cluster + logwealthshadow$|$Cluster;

HC0300 $<-$ 1$|$Cluster + logwealthshadow$|$Cluster;

HC0400 $<-$ 1$|$Cluster + logwealthshadow$|$Cluster;

HD0100 $<-$ 1$|$Cluster + logwealthshadow$|$Cluster;

   HD1100 $<-$ 1$|$Cluster + logwealthshadow$|$Cluster;
   
   HD1200 $<-$ 1$|$Cluster + logwealthshadow$|$Cluster;
   
   HD1300 $<-$ 1$|$Cluster + logwealthshadow$|$Cluster;
   
   HD1400 $<-$ 1$|$Cluster + logwealthshadow$|$Cluster;
   
   HD1500 $<-$ 1$|$Cluster + logwealthshadow$|$Cluster;
   
logwealth $<-$ 1$|$Cluster;

logwealth$|$Cluster;}
\end{document}